\documentclass[noinfoline]{imsart}

\usepackage{amsfonts,amssymb,amsmath,amscd,amsthm,latexsym}
\usepackage{natbib}
\usepackage[]{units}

\usepackage[french]{babel}
\usepackage[utf8]{inputenc}

\newcommand\CB{\mathcal{B}}
\renewcommand\CD{\mathcal{D}}

\newcommand\CN{\mathcal{N}}

\newcommand\oh{\nicefrac{1}{2}}
\newcommand\dd{\text{d}}
\newcommand\bx{\mathbf{x}}

\begin{document}

\begin{frontmatter}

\title{Des spécificités de l'approche bayésienne et de ses justifications en statistique inférentielle\protect\thanksref{T1}}
\runtitle{Sur les spécificités de l'approche bayésienne en statistique inférentielle}
\thankstext{T1}{Christian P. Robert, CEREMADE, Universit{\' e} Paris-Dauphine, 75775 Paris cedex 16, France
{\sf xian@ceremade.dauphine.fr} et Department of Statistics, University of Warwick, Coventry CV4 7AL, United Kingdom. 
Recherche financée en partie par l'Agence Nationale de la Recherche (ANR,
212, rue de Bercy 75012 Paris) sur le contrat 2012--2015 ANR-11-BS01-0010 ``Calibration'' et par une chaire
senior de l'Institut Universitaire de France.}

\begin{aug}
 \author{\snm{{\sc Christian P.~Robert}}}
 \affiliation{Universit{\'e} Paris-Dauphine, CEREMADE, University of Warwick, Department of Statistics,
 et CREST, Paris}
\end{aug}

\begin{abstract}
This book chapter (written in French) is a review of the foundations of the Bayesian approach to statistical
inference, relating to its historical roots and some philosophical arguments, as well as a short presentation
of its practical implementation.

Ce chapitre d'un ouvrage de philosophie des sciences portant sur les méthodes bayésiennes, à paraître
aux éditions Matériologiques, vise à donner les fondements de l'approche bayésienne en statistique inférentielle,
ses racines historiques et ses justifications philosophiques, ainsi qu'à présenter des illustrations
de sa mise en {\oe}uvre pratique.
\end{abstract}

\end{frontmatter}

\section{Introduction}

La statistique bayésienne est une approche spécifique de la statistique
in\-fé\-ren\-tiel\-le qui propose une réponse à la fois unitaire et globale au problème
inférentiel, dans le cadre paramétrique comme dans le cadre non-paramétrique.
Elle se distingue des approches dites ``classiques" par la construction et
l'utilisation d'une loi de probabilité sur l'ensemble des notions pouvant faire
l'objet d'une inférence. On peut légitimement se demander pourquoi la
distinction entre cette approche bayésienne et les méthodologies (plus)
classiques est nécessaire--et pourquoi elle n'est pas ``classique"--, d'autant
qu'elle s'accompagne de débats philosophiques particulièrement virulents et de
positions militantes pouvant parfois évoquer des dérives sectaires, attitudes
qui ne se retrouvent pas dans les autres approches de la statistique
inférentielle. 

Le principe de l'inférence bayésienne se résume assez simplement : étant donné
un modèle statistique, l'ensemble des constituants (paramètres et/ou fonctions)
inconnus de ce modèle est traité comme une variable aléatoire---potentiellement
de dimension infinie---, donc munie d'une loi de probabilité, et l'ensemble des
réponses inférentielles se fonde sur la loi de cette variable aléatoire,
conditionnellement aux données. Ce qui fait la beauté de cette approche et
explique en partie son attractivité est que la démarche inférentielle est alors
quasi-automatique, étant donné cette loi et une mesure des performances des
procédures dite {\em fonction de coût}. Dans les approches alternatives, la
seule variable aléatoire est celle correspondant aux données et la contruction
des procédures inférentielles est généralement ouverte (entre estimateurs des
moments, du maximum de vraisemblance, etc.) La contrepartie à ce caractère
automatique, que certains pourraient qualifier de beauté vénéneuse!, est que la
modélisation des inconnues en variable aléatoire exige un choix de loi de
probabilité, choix subjectif opéré par le statisticien bayésien car ne reposant
pas directement sur l'observation. Cet aspect de l'approche bayésienne
concentre la majorité des critiques de la communauté statistique, une critique
secondaire (et plus répandue dans la communauté du {\em machine learning})
étant l'obligation faite au statisticien de construire intégralement le modèle
statistique sur les données, ce qui empêcherait le traitement de structures
complexes et/ou  de grandes tailles.

Ce chapitre présente brièvement le développement historique de l'approche bayésienne,
depuis Bayes et Laplace jusqu'à nos jours, ainsi que les motivations
philosophiques et méthodologiques qui la sous-tendent.  La seconde partie
présente quelques éléments de mise en \oe uvre au travers d'exemples simples.
Nous dirigeons le lecteur vers \cite{robert:2005} pour une couverture plus
complète (en fran\c cais) de cette méthodologie spécifique, de nombreux autres
ouvrages étant disponibles en anglais.

Quelques avertissements au lecteur sont de mise quant au contenu de ce chapitre
: ne pouvant y définir correctement les concepts de probabilité nécessaires, je suppose que
les lecteurs sont suffisamment familiers avec ceux-ci, à un niveau de fin de
Licence (3ième année d'université), pour manipuler les concepts de probabilité
conditionnelles. Par ailleurs, ces mêmes lecteurs devront se référer à leur
source favorite en ce qui concerne les distributions usuelles (Wikipédia, bien
sûr, ou l'annexe de \citealp{robert:2005}).

\section{Un peu d'histoire sur le développement du raisonnement bayésien, des origines
à la quasi-extinction post-kolmogorienne, et au renouvellement modélisateur}

\subsection{Des probabilités inverses comme définition de la statistique}

Le concept de ``statistique bayésienne" part du néologisme ``bayésien", tiré du nom de Thomas
Bayes,\footnote{Thomas Bayes (1702?-1761) était un prêtre presbytérien non-conformiste,
membre de la Royal Society, qui publia à titre posthume un {\em Essay towards solving a Problem 
in the Doctrine of Chances}. On connaît très peu de détails sur sa vie et le seul portrait de lui dont
on dispose demeure incertain. Voir \cite{mcgrayne:2011} pour une introduction et \cite{dale:1999}
pour une étude plus profonde sur les contributions de Bayes à la théorie qui porte à présent son
nom.} qui introduisit le théorème qui porte à présent son nom dans un article posthume de 1763,
il y a 250 ans. Ce théorème exprime une probabilité conditionnelle en termes de
la probabilité conditionnelle inverse pondérée par les probabilités marginales,
$$
P(A|B) = \dfrac{P(B|A) P(A)}{P(B)}\,,
$$
ce qui a valu à la statistique (alors uniquement envisagée sous cet angle
bayésien) d'être appelée ``probabilités inverses" pendant plus d'un siècle, de
\cite{laplace:1812} à \cite{keynes:1921}, avant que Fisher\footnote{Ronald
Fisher, statisticien et généticien anglais, est à l'origine de la notion de
vraisemblance. Féroce critique des approches alternatives à la sienne, et en
particulier de la perspective bayésienne, il proposa à la fin de sa carrière la
notion de statistique fiducière qui, s'appuyant sur des quantités pivotales,
ressemblait formellement à une modélisation bayésienne non-informative.}
n'introduise le terme ``bayésien" \citep{fienberg:2006}.\footnote{Pierre Simon
de Laplace a contribué à formaliser et à généraliser la mise en \oe uvre des
probabilités inverses, bien plus que Thomas Bayes. Cette approche aurait donc
mérité de s'appeler laplacienne plutôt que bayésienne. Notons également que, si
Keynes a été formé à la statistique suivant des principes bayésiens, il rédige
son traité de 1921 dans un esprit assez critique, sans pour autant proposer une
alternative constructive \citep{robert:2010}.} Bien que la formule (ou
théorème) de Bayes soit une conséquence directe de la définition des
probabilités (et densités) conditionnelles, son application à des
problématiques statistiques, où une observation $x$ dépend d'un paramètre
inconnu $\theta$ est effectivement appropriée, au sens où le contexte {\em
inverse} ce qui est connu et ce qui est inconnu. Effectuer l'inversion pour
obtenir l'information contenue dans $x$ à propos de $\theta$ conduit à définir
la loi {\em  a posteriori}, loi conditionnelle de $\theta$ sachant $x$.

\subsection{Des notions d'a priori et d'a posteriori, et sur l'illusion des paramètres aléatoires}\label{sec:p!p}

Pour que la loi a posteriori soit définie, le modèle doit non seulement
comprendre une loi des observations, de densité $f(x;\theta)$ ou $f(x|\theta)$
mais également une loi de probabilité sur le paramètre $\theta$, de densité
$\pi(\theta)$ et appelée loi a priori. Dans ce cas, la loi a posteriori
s'obtient par la formule de Bayes,
$$
\pi(\theta|x) = \dfrac{f(x|\theta)\pi(\theta)}{\int_\Theta f(x|\theta)\pi(\theta)\,\text{d}\theta}\,.
$$
Cette formule, présente dans l'essai de Thomas Bayes de 1763, est la version
générale du théorème d'inversion ci-dessus. Le dénominateur de la fraction est
à la fois la densité marginale associée à l'observation $x$ et la constante de
normalisation permettant de transformer le numérateur en densité de
probabilité.  Dans ce contexte, où $\theta$ est traité comme une variable
aléatoire, la loi des observations apparaît comme une loi conditionnelle à la
valeur du paramètre, plutôt que comme une loi indicée par $\theta$ comme dans
l'approche classique.

Cette distinction entre paramètre inconnu (mais fixe) et paramètre aléatoire
peut à la fois paraître fondamentale et sembler condamner l'approche bayésienne
comme inappropriée vis-à-vis de la compréhension moderne---élaborée par
Kolmogorov et ses successeurs---de la notion de probabilité. La distinction
faite entre probabilité comme stabilisation des fréquences (loi des grands
nombres) et probabilité comme quantification d'un degré d'incertitude met en
lumière l'imprécision et la subjectivité liées à la seconde approche.  C'est
certainement la perspective qu'adopta Fisher très rapidement dans sa carrière
et la raison de sa querelle avec Jeffreys au cours des années 1930
\citep{robert:chopin:rousseau:2009}. Bien que fondée elle aussi sur la notion
de vraisemblance,
$$
\ell(\theta|x) = f(x|\theta)\,,
$$
qui reflète également un principe d'inversion, l'approche de Fisher refuse
l'interprétation de $f(x|\theta)$ comme densité conditionnelle et la
probabilisation de $\theta$. Il est cependant abusif de faire de cet aspect de
l'analyse bayésienne autre chose qu'une source de discussion philosophique. En
effet, le passage de la notion de paramètre {\em inconnu} à la notion de
paramètre {\em aléatoire} est incompatible avec la plupart des expériences en
particulier dans les sciences physiques. La modélisation statistique suppose au
contraire l'existence d'un paramètre fixe $\theta$, sur lequel elle vise à
obtenir une information aussi précise que possible.  La constante de Hubble, la
vitesse de la lumière, le coefficient de diffraction d'un prisme sont autant
d'exemples où le concept d'aléa sur le paramètre ne fait pas sens. Cependant, ce que propose
l'approche bayésienne se situe à un autre niveau sémantique  qui ne remet pas en cause
cette modélisation de la réalité. La loi a posteriori est utilisée comme un
nouvel (et efficace) outil de résumé de l'information disponible sur $\theta$,
sans remettre en cause l'existence de ce paramètre inconnu et
{\em non aléatoire}. En d'autres termes, la loi a posteriori est un outil
de représentation de l'information disponible sur $\theta$ une fois les
observations obtenues, elle reflète l'incertitude inhérente aux données et
à l'expérimentateur plutôt qu'un aléa physique sur ce paramètre. 

Cette distinction entre outil de représentation et véritable aléa a cependant
presque causé l'extinction définitive de l'approche bayésienne lorsque, au
tournant du siècle, et comme dans de nombreuses autres branches des
mathématiques, Kolmogorov et ses collègues probabilistes ont formalisé la
théorie des probabilités. Il devenait alors difficile de donner un sens
mathématique à la loi a priori si elle ne correspond pas à un véritable
phénomène aléatoire mais plutôt à une traduction de ce que l'expérimentateur
est prêt à parier sur les valeurs possibles du paramètre.\footnote{C'est
l'argument du {\em Dutch Book}, voir \cite{berger:1985}.}

Cet aspect subjectif de la loi a priori, résultant du choix de
l'expérimentateur, peut apparaître comme un élémént réducteur de la perspective
bayésienne, mais il permet de refléter les informations (et leur degré
d'imprécision) dont dispose cet expérimentateur et donc conduit à une inférence
plus précise et plus riche, de ce fait.  Que deux expérimentateurs adoptent
deux lois a priori différentes ne devrait pas plus porter à critique que le
fait que deux expériences (donc deux séries d'observations différentes)
conduisent à des fonctions de vraisemblance différentes. Ni que deux statisticiens
``classiques" choisissent l'une un estimateur du maximum de vraisemblance et l'autre
un estimateur des moments. Il est fondamental
d'observer que l'approche bayésienne ne dispose pas d'une {\em seule loi a
priori}, une sorte de Graal qu'il conviendrait d'obtenir après de longues
recherches! Cette vision, souvent observée dans la littérature appliquée, est du
même ordre que celle qui attribue au paramètre $\theta$ un véritable caractère
aléatoire. (Bien entendu, dans certains contextes comme ceux des modèles à
effets aléatoires ou de la prévision, certaines composantes de $\theta$ sont
effectivement considérées comme aléatoires, mais elles le sont aussi dans
l'approche classique de ces modèles.) Dans l'approche bayésienne, comme dans
dans la plupart des autres écoles de statistique, parler de la ``vraie" valeur
du paramètre fait sens.

Comme nous le verrons dans les sections suivantes,
disposer d'une loi a posteriori permet de construire des procédures
inférentielles (tests, intervalles de confiance, densités prédictives) 
de la manière la plus naturelle possible, ce qui explique entre
autres la persistance de cette approche, contre vents et marées
\citep{mcgrayne:2011}, depuis 250 ans. 

\subsection{Une justification par le principe de vraisemblance}

Tandis que la vision de Fisher, fondée sur la vraisemblance, n'admet pas une
extension vers une modélisation bayésienne, il existe un principe de vraisemblance,
formalisé par \cite{birnbaum:1962}, dont découle comme implémentation première la méthodologie
bayésienne. Bien que ce principe soit régulièrement remis en cause \citep{mayo:2010}, il 
sous-tend suffisamment cette approche pour que nous le rappelions brièvement ici.

Le {\em principe de vraisemblance} impose à l'expérimentateur confronté à deux
séries d'observations $\bx_1$ et $\bx_2$ portant sur le même paramètre $\theta$
ayant des fonctions de vraisemblance proportionnelles, donc telles que, pour
tout $\theta$,
$$
\ell(\theta|\bx_1) \propto \ell(\theta|\bx_2)\,, 
$$
de conduire à la même inférence sur ce paramètre $\theta$. C'est naturellement le cas si
l'expérimentateur adopte une démarche bayésienne puisque les lois a posteriori sont
alors identiques. Il en va de même pour l'estimation par maximum de vraisemblance. 
L'intuition derrière le principe est que l'information apportée
par les deux échantillons sur le paramètre $\theta$ est la même. 
Cette situation de fonctions de vraisemblance proportionnelles se
produit par exemple lorsqu'on compare un échantillonnage binomial,
$m\sim\mathcal{B}in(n,p)$, à un échantillonnage binomial négatif,
$n\sim\mathcal{N}eg(m,p)$, ce qui correspond à un sondage avec un nombre fixe
de personnes interrogées contre un sondage avec un nombre fixe de personnes
avec une réponse donnée. Le nombre d'essais total $n$ et le nombre de succès
$m$ sont les mêmes, mais l'aléa portant sur deux parties différentes, les
fonctions de vraisemblance sont proportionnelles,
\begin{align*}
\ell_1(p|n) &= {n \choose m} p^m (1-p)^{n_m} \\
	&\propto {n-1 \choose m-1} p^m (1-p)^{n_m} = \ell_2(p|m)\,.
\end{align*}
Bien que cet exemple puisse paraître très artificiel, il sous-tend la classe des 
problèmes de régles d'arrêt, où un échantillon $(x_1,\ldots,x_n)$ est de taille $n$
aléatoire, la loi de $n$ ne dépendant pas du paramètre $\theta$. 
Une inférence distinguant un échantillon $(x_1,\ldots,x_n)$ iid\footnote{Abréviation
de {\em indépendants et identiquement distribués}.} d'un
échantillon $(x_1,\ldots,x_n)$ produit par une règle d'arrêt contredit le principe de
vraisemblance. Plus généralement, des méthodologies fondées sur des propriétés fréquentistes
comme les $p$-values ne s'accordent pas avec ce principe. 

\cite{birnbaum:1962} a démontré que le principe de vraisemblance découlait
logiquement de la conjonction de deux autres principes, le principe de conditionalité
et le principe d'exhaustivité, à savoir que, (i) si deux expériences sont
possibles pour mesurer $\theta$ et que l'une est choisie au hasard, seule compte l'expérience effectivement
réalisée, et que (ii) l'inférence ne doit dépendre des données que via des statistiques
exhaustives.\footnote{La réfutation de cette démonstration par
\cite{mayo:2010} semble découler d'une définition tautologique de la notion d'inférence.}  
Comme le discutent \cite{berger:wolpert:1988}, la mise en \oe
uvre du principe de vraisemblance conduit naturellement à une construction
bayésienne (qui ne dépend effectivement que de la fonction de vraisemblance),
même s'il existe des méthodologies alternatives (comme le test du rapport de
vraisemblance) respectant le principe de vraisemblance et couvrant certains
aspects de l'inférence. 

\section{Sur la construction des lois a priori, sur leur calibration et des
réponses aux critiques afférentes}

Comme signalé ci-dessus, la loi a priori est choisie par l'expérimentateur et
témoigne à un degré ou un autre d'un choix (comme peut l'être la sélection de
la procédure inférentielle voire de la loi des observations). Ces lois sont
souvent choisies dans des classes de lois standard pour faciliter leur
utilisation, même après l'avénement de techniques de simulation plus
puissantes.  L'impact de ce choix sur la réponse inférentielle est non-nulle,
même s'il disparaît à mesure que la taille de l'échantillon grandit, du fait
de la consistance--convergence de l'estimation vers la vraie valeur du paramètre quand
la taille de l'échantillon tend vers l'infini--de l'approche bayésienne dans un grand nombre de situations.
Il peut être évalué de manière analytique ou numérique, mais aussi comparé à
des solutions dites de référence \citep{jeffreys:1939,berger:bernardo:1992},
décrites ci-dessous.

\subsection{Des lois conjuguées comme éléments de base de la modélisation a priori}

Dans le cadre de lois d'observations standard comme la loi normale, il existe
des familles de lois a priori facilitant le calcul de la loi a posteriori. Bien
que leur motivation soit surtout fondée sur le fait qu'elles simplifient de la
dérivation de la loi a posteriori (plutôt que
sur une caractéristique particulière de l'information a priori), elles n'en
constituent pas moins un élément de base dans la construction effective de lois
a priori plus personnelles. Par ailleurs, et d'un point de vue beaucoup plus
pratique elles sont l'élément constitutif du
logiciel BUGS \citep{lunn:thomas:best:spiegelhalter:2000,lunn:bugs:2012}, principal
outil de l'analyse bayésienne de données car
ces lois conjuguées permettent au logiciel de construire automatiquement l'algorithme correspondant
de Monte Carlo par échantillonnage de Gibbs.\footnote{Nous rappelons d'une part que {\em BUGS} signifie {\em
Bayesian analysis Using Gibbs Sampling} et d'autre part que le physicien Josiah
Willard Gibbs n'a rien à voir avec les algorithmes de simulation
conditionnelle qui portent à présent son nom.  Ces algorithmes ont simplement
été utilisés pour la première fois sur des champs de Gibbs
\citep{geman:geman:1984}. Ils constituent un exemple d'algorithmes de Monte Carlo par
chaînes de Markov et ont énormément contribué à l'explosion des applications bayésiennes
au début des années 1990 \citep{robert:casella:2004}.} Ces lois sont dites conjuguées (à
une famille de fonctions de vraisemblance).

Les lois conjuguées sont intrinséquement associées aux lois de familles exponentielles:
$$
f(x|\theta) = h(x)\,\exp\{ \theta\cdot R(x) - \psi(\theta) \}\,,
$$
où $\theta$ est un paramètre de dimension $d$ et $R$ une fonction à valeurs dans $\mathbb{R}^d$, $\psi(\theta)$
permettant la normalisation de la densité de probabilité. Ainsi, ces densités sont associées aux lois a
priori de la forme
$$
\pi(\theta) \propto \exp\{ \theta\cdot \rho - \lambda\psi(\theta) \}\,,
$$
où $\lambda>0$ et $\rho\in\mathbb{R}^d$ sont contraints par l'intégrabilité de la fonction,
puisque
$$
\pi(\theta|x) \propto \exp\{ \theta\cdot [\rho+R(x)] - [\lambda+1]\psi(\theta) \}\,.
$$
La loi a posteriori est donc définie par un changement de paramètres, de
$\lambda$ à $\lambda+1$, et de $\rho$ à $\rho+R(x)$. 

Un point crucial est que ces lois conjuguéees ne peuvent exister {\em que} pour
les vraisemblances découlant des familles exponentielles, en vertu du Lemme de
Pitman-Koopman sur l'existence de statistiques exhaustives (trop technique pour
être détaillé ici, voir par exemple \citealp{robert:2005}). Elles sont cependant
disponibles pour un grand nombre de lois classiques, comme la loi normale à un
ou deux paramètres, les lois binomiale, binomiale négative, de Poisson, gamma à
un ou deux paramètres, Dirichlet, de Wishart... Par exemple, pour des
observations $x_1,\ldots,x_n$ normales de loi $\mathcal{N}(\theta,\sigma^2)$,
où $\sigma$ est connu, la loi conjuguée est aussi normale
$\theta\sim\mathcal{N}(\mu,\tau^2)$. En effet, la loi de $\theta$ a posteriori
est la loi normale
$$
\theta|x_1,\ldots,x_n \sim \mathcal{N}(\{\tau^{-2}+n\sigma^{-2}\}^{-1}\{\tau^{-2}\mu
+n\sigma^{-2}\bar x_n\},\{\tau^{-2}+n\sigma^{-2}\}^{-1})\,.
$$

Les lois conjuguées ayant une forme {\em imposée} par la loi des observations, elles
ne sont pas à même de refléter toute sorte d'information a priori, tout en
demandant la spécification des hyperparamètres $\lambda$ et $\rho$. Une
extension efficace et robuste de ces lois est fournie par les mélanges, discrets ou
continus, obtenus par l'introduction de lois sur les paramètres $\lambda$ et
$\rho$, en particulier parce qu'elles autorisent à leur tour des
implémentations par l'algorithme de Gibbs \citep{diaconis:ylvisaker:1979}.

\subsection{Sur les lois de Jeffreys et de l'impossibilité de trouver la loi la moins informative}

Dans une situation où l'information a priori n'est pas disponible, et où l'expé\-rimen\-ta\-teur renacle
à faire un choix, il est tentant de chercher à proposer une loi reflétant ce manque d'information.
Hélas (ou pas!), il n'existe pas de loi ``la moins informative" et on peut seulement, au mieux, 
définir une procédure automatique de construction d'une loi a priori de référence, à l'aune de
laquelle les lois a priori subjectives peuvent être évaluées. Il s'agit donc d'une convention et
non d'une optimalité quelconque au titre de la faible information a priori. Même si ces lois 
de référence sont
souvent appelées non-informatives, elles contiennent néanmoins des informations sur le paramètre
et ne peuvent prétendre représenter une complète ignorance \citep{kass:wasserman:1996}.

Une illustration de cette impossibilité est fournie par le principe de la raison insuffisante
de \cite{laplace:1812}: par extension du cas d'un ensemble fini de paramètres, Laplace propose
d'utiliser une loi uniforme dans toute situation non-informative. Si l'espace des paramètres
n'est pas compact, cela exige l'emploi d'une mesure de Lebesgue en lieu et place d'une loi de
probabilité, ce qui ne pose pas problème en soi du moment que la loi a posteriori est définie, et
surtout la solution dépend de la paramétrisation adoptée, puisque la loi uniforme ne ``résiste" pas à
un changement de variable. Bien que cette solution reste celle adoptée tout au long du 19ième siècle, elle
cristallise les critiques naissantes sur le paradigme bayésien.

Il faut en fait attendre les années 1930 et Harold Jeffreys\footnote{Harold Jeffreys fut à la fois
mathématicien, statisticien, géophysicien et astronome. Son livre, {\em Theory of Probability},
demeure une référence sur la formalisation de l'approche bayésienne, écrite à une époque où celle-ci
n'était plus si populaire, même si certains aspects du livre sont criticables sur le plan mathématique
\citep{robert:chopin:rousseau:2009}.} pour voir apparaître une perspective globale sur le choix des
lois de référence. Ironiquement, cette solution emprunte à Fisher en ce qu'elle est fondée sur
l'information du même nom. Si l'information de Fisher\footnote{Cette matrice indexée par
le paramètre mesure le pouvoir des données à discriminer entre deux valeurs du paramètre. 
Elle donne une représentation de la courbure de la surface de vraisemblance et apparaît via
son inverse dans la variance asymptotique des estimateurs standard.}
associée à un modèle est définie par
$$
I(\theta) = \mathbb{E}_{\theta}\left[{\partial\ell\over\partial\theta^t}\;
              {\partial\ell\over\partial\theta}\right]
$$
la loi de Jeffreys associée est donnée par
$$
  \pi^\ast (\theta ) \propto |I(\theta)|^{1/2}
$$
soit donc la racine du déterminant de l'information. Plusieurs remarques sont de mise :
\begin{enumerate}    
\item  la solution de Jeffreys est invariante par changement de paramétrisation du modèle, 
grâce à la formule du jacobien ;
\item  la loi est directement dépendante de l'information, ce qui signifie que les régions où
l'information est plus importante sont privilégiées, car les données y sont plus discriminantes ;
\item  cette construction est en général compatibles avec l'utilisation de lois invariantes 
sous l'action d'un groupe (comme celui des translations), les mesures de Haar ;
\item cette construction ne garantit en rien l'intégrabilité de $\pi^\ast$, ce qui signifie que
les lois de Jeffreys seront souvent des mesures ;
\item  les lois de Jeffreys dépendent de l'intégralité de la loi des observations, donc ne
respectent pas le principe de vraisemblance. 
\end{enumerate}
Une illustration de ce dernier point est fournie par l'opposition entre la loi binomiale,
$\CB(n,\theta)$ pour qui la loi de Jeffreys est la loi $\CB e(1/2,1/2)$, et la loi binomiale
négative, $\CN eg(x,\theta)$, pour qui la loi de Jeffreys est la mesure (impropre, c'est à dire
ne pouvant être normalisée en une loi de probabilité) de densité $1/\theta \sqrt{1-\theta}$.

Comme cette construction universelle de lois de référence peut donner naissance
à des ``monstres", par exemple dans le cas de l'estimation de $||\theta||^2$
quand $x\sim\CN(\theta,I_p)$ et $p$ grand, il existe des déterminations de
lois de référence qui distinguent entre paramètres d'intêret et paramètres de
nuisance avant de construire des lois de Jeffreys conditionnelles et marginales
sur les deux groupes.  Nous renvoyons le lecteur à \cite{berger:bernardo:1992}
et \cite{bernardo:smith:1994} pour des entrées sur cette extension qui demeure
peu utilisée à ce jour. Voir aussi \cite{kass:wasserman:1996} pour une revue de
certains principes de détermination des lois ``non-informatives". En lien avec
ces principes, insistons une nouvelle fois sur le fait que, de même qu'il
n'existe pas ``une" loi a priori unique qu'il faudrait découvrir, il n'existe
pas non plus ``une" seule loi a priori non-informative. Il s'agit bien de poser
une référence, qui peut servir à la fois à analyser des problèmes sans information
visible et à comparer différentes lois a priori, si besoin. (Cette référence se trouve faire
défaut dans le cadre des tests, à moins d'accepter quelques compromis avec le paradigme
bayésien, \citealp{robert:2005}.)

\subsection{Sur l'utilisation de lois en dimension infinie et de l'apparition de restaurants ethniques}

Nous avons mentionné dans l'Introduction que l'approche bayésienne permettait
également de traiter de problèmes dans un cadre non-paramétrique. Cela signifie
par exemple que, à partir d'un échantillon iid $x_1,\ldots,x_n$ de loi
inconnue\footnote{Notons que cette hypothèse est peu restrictive en ce qu'elle
correspond {\em in fine} à celle d'échangeabilité sur la loi du $n$-uplet,
suivant la représentation de Bruno de Finetti. Voir par exemple
\cite{bernardo:smith:1994}.} $F$, une loi a posteriori peut être construite sur
$F$. Les lois a priori qui sous-tendent ces constructions sont donc des lois
sur des espaces de fonctions comme l'ensemble des densités de probabilité. 
Par exemple, la loi de Dirichlet {\em fonctionnelle},
$\mathcal{D}ir(\alpha,G_0)$, est définie à partir d'un coefficient de
précision $\alpha>0$ et d'une moyenne a priori $G_0$, loi de probabilité. Cette
loi \citep{ferguson:1974} généralise la loi de Dirichlet sur le simplexe $\mathcal{D}ir_d
(\beta_1,\ldots,\beta_d)$ au sens où, pour toute partition $\{A_1,\ldots,A_d\}$
de l'espace des observations, si $G\sim\mathcal{D}ir(\alpha,G_0)$, alors
$$
(F(A_1),\ldots,F(A_d)) \sim \mathcal{D}ir_d (\alpha G_0(A_1),\ldots,\alpha G_0(A_d))\,.
$$
La loi a posteriori correspondante est conjuguée,
$$
F|x_1,\ldots,x_n \sim \mathcal{D}ir(\alpha+n,G_0+\hat F_n)\,,
$$
où $\hat F_n$ dénote la distribution empirique associée à l'échantillon $x_1,\ldots,x_n$, ce
qui signifie que la variable aléatoire n'est pas absolument continue par rapport à la mesure
de Lebesgue, même si $G_0$ l'est. En particulier,
si $F\sim \CD(\alpha,G_0)$, la loi conditionnelle de $x_1$ sachant $(x_2,\ldots,x_n)$ est
de la forme
$$
\frac{\alpha}{\alpha+n-1} F_0 + \frac{1}{\alpha+n-1} \sum_{i=2}^n \delta_{x_i}\,.
$$
L'évaluation complète de la loi a posteriori de $F$ exige l'emploi d'outils de simulation
qui ne seront pas abordés ici (voir par exemple \citealp{denison:holmes:mallick:smith:2002}
et \citealp{hjort:holmes:mueller:walker:2010}).

Une représentation équivalente du processus de Dirichlet est appelée {\em
processus du restaurant chinois} pour la raison suivante : générer un
échantillon de taille $n$ suivant le processus de Dirichlet est équivalent à
simuler un flux d'arrivées de clients dans un restaurant chinois stylisé.
Chaque client de ce restaurant s'assied à une table déjà occupée avec une
probabilité en proportion du nombre de clients présents à cette table et à une
nouvelle table avec probabilité proportionnelle à $\alpha$. D'autres versions
du processus de Dirichlet sont fondées sur des successions de partitions de
l'intervalle $[0,1]$ appelées ``stick breaking" en raison de leur construction
récurrente \citep{sethuraman:1994}. Ces différentes interprétations sont
utilisées soit au niveau théorique, pour valider la convergence des estimateurs
résultants \citep{vandervaart:1998,barron:1999,ghosal:etal:2008}, soit au
niveau computationnel, pour simuler ces processus
\citep{hjort:holmes:mueller:walker:2010,mueller:mitra:2013}.

De nombreuses extensions de cette loi a priori existent dans la littérature,
fondées sur différents processus comme les processus gaussiens ou le processus
de Lévy. Des représentations par mélanges permettent en particulier de dépasser
le défaut des processus de Dirichlet de ne simuler que des lois à support
discret.  Une extension de la représentation du restaurant chinois s'appelle le
buffet indien (!) et permet le mélange de plusieurs caractéristiques au lieu
d'imposer une partition \citep{griffiths:ghahramani:2006}.

Enfin, notons que les outils traditionnels de modélisation mathématique d'espaces de
fonctions, comme par exemple l'emploi de bases d'ondelettes, autorisent une autre
forme de représentation bayésienne de l'inférence fonctionnelle
\citep{vidakovic:1999, clyde:george:2000}.

\section{De la mise en \oe uvre des principes bayésiens}

Nous décrivons dans cette partie quelques mises en \oe uvre de l'inférence bayésienne dans
des modèles standard. Il s'agit en grande partie d'illustrations et nous renvoyons le lecteur
par exemple à \cite{robert:2005} pour une perspective plus complète.

\subsection{De la loi a posteriori comme pivot de l'inférence}

Si la loi a priori $\pi(\theta)$ est l'élément central qui concentre
l'essentiel des critiques sur la perspective bayésienne, la loi a posteriori
$\pi(\theta|x)$ est l'élément central permettant de conduire l'inférence
bayésienne. Une fois cette loi construite, et le modèle accepté (voir
ci-dessous), il n'est plus nécessaire\footnote{Ce résultat découle aussi
directement du principe de vraisemblance.} de conserver les données !
En effet, les diverses procédures inférentielles sont toutes construites
(automatiquement) à partir de cette loi : estimateurs, variations, intervalles
de confiance, tests, choix de variables, choix de modèle, prévisions, etc.
Certains défendent en fait la définition de la loi a posteriori comme ultime
but de l'inférence bayésienne, considérant que les procédures ci-dessus ne sont
que des résumés qui dégradent le contenu informatif de la loi a posteriori.
Bien qu'une loi de probabilité ne puisse en effet se résumer à certains de ses
moments, il est néanmoins pertinent de fournir des réponses aux questions des
clients et des décideurs, pour lesquels la loi a posteriori demeure un concept
abstrait. Cette perspective pragmatique est particulièrement pertinente dans
les problèmes à paramètres de grande dimension, qui pour la plupart correspondent à
des paramètres de nuisance.\footnote{Répétons ici la remarque que la prise en
compte de la distinction entre paramètres d'intérêt et paramètres de nuisance a
conduit \cite{berger:bernardo:1992} à proposer des lois de référence
reproduisant cette distinction et éliminant les paramètres de nuisance par une
intégration suivant une loi de Jeffreys conditionnelle.}

La loi a posteriori $\pi(\theta|x)$ étant construite, on peut dériver (analytiquement
ou non) les estimateurs ponctuels que sont la moyenne et la médiane a posteriori de
n'importe quelle transformation $\varphi(\theta)$, ainsi que les évaluations de
leur variabilité fournies par les écart-types a posteriori. De plus, la loi a
posteriori induit des lois a posteriori marginales pour toute transformation
$\varphi(\theta)$, déduites par projection du modèle probabiliste joint,
$\varphi\sim\pi(\varphi|x)$. De ces lois marginales peuvent se déduire des
intervalles ou des régions de confiance naturelles, les régions de plus forte
densité a posteriori (ou régions HPD pour {\em highest posterior density})
$$
\mathcal{C}(x)  = \{ \varphi;\ \pi(\varphi|x) > \tau \}\,,
$$
dont la borne $\tau$ peut se calibrer en fonction du taux de couverture
$\alpha$ souhaité pour cette région. (Elle dépendra alors de $x$.)
Contrairement aux intervalles fondés sur l'approximation normale des ``deux
sigma", ces régions ont une couverture exacte et peuvent être assymétriques.
Elles sont par contre dépendantes de la paramétrisation choisie (ou, ce qui est
équivalent, de la mesure utilisée pour mesurer les volumes, voir \citealp{druilhet:marin:2007}).
Il est même possible de discuter de la vraisemblance d'une hypothèse comme $H_0:\,\varphi=0$
en examinant si la valeur $\varphi=0$ appartient à la région HPD, bien que cette perspective
ne soit pas conseillée puisqu'elle ne prend pas en compte l'action résultant d'un rejet
de $H_0$.

Comme simple illustration, prenons l'exemple d'un phénomène binaire, à valeurs
dans $\{0,1\}$, observé durant $n$ répétitions indépendantes et identiquement
distribuées (i.i.d.). Le nombre de $1$, $X$, est alors modélisé par une loi
binomiale $\mathfrak{B}(n,p)$ où $p$ est la probabilité d'obtenir une valeur
$1$. Si on associe à ce paramètre $p$ une loi de Jeffreys, 
$$
\pi(p) = 1 \big/ \sqrt{p(1-p)}\,,
$$
qui correspond à une loi Béta $\mathfrak{B}e(\oh,\oh)$ une fois normalisée,
la construction de la loi a posteriori est immédiate :
\begin{align*}
\pi(p|x)&\propto \pi(p) \times \ell(p|x) \\
	&\propto p^x (1-p)^{n-x} \big/ \sqrt{p(1-p)} \\
  	&\propto p^{x-\oh} (1-p)^{n-x-\oh}\,,
\end{align*}
qui correspond à une loi Béta $\mathfrak{B}e(x+\oh,n-x+\oh)$ (toujours après normalisation). 
Si, par exemple, $n=232$ et $x=117$, la loi a posteriori est la loi Béta $\mathfrak{B}e(117.5,115.5)$,
de mode {\em et} moyenne $117.5/233=0.504$. L'intervalle de crédibilité à queues égales, valant $[0.440,0.558]$
pour un niveau de crédibilité $\alpha=0.95$, est plus facile à calculer que la région HPD.\footnote{Une résolution
numérique plus poussée de cette dernière conduit à... $[0.440,0.558]$, soit donc le même intervalle (pour cette
précision) que la solution symétrique!} Si à présent nous cherchons à tester l'hypothèse nulle et ponctuelle $H_0:\, 
p=\oh$, le facteur de Bayes en faveur de $H_0$ (voir Section 4.3) est donné par
$$
\mathfrak{B}_{01}(x) = \oh^n \Big/ \dfrac{B(x+\oh,n-x+\oh) }{ B(\oh,\oh)}\,,
$$
où $B(\alpha,\beta)$ dénote la constante de normalisation de la densité de la loi $\mathfrak{B}e(\alpha,\beta)$.
Dans le même exemple numérique que ci-dessus, ce facteur de Bayes vaut $\pi(\oh|x)/\pi(\oh)=18.95$, donc exprime 
un soutien assez important en faveur de $H_0$.

Même si les estimateurs ci-dessus sont des produits de la loi a posteriori, il est
légitime de se demander comment les comparer et quel estimateur choisir comme ``meilleur
résumé". \`A l'exception de cas spécifiques, il n'existe pas de réponse à cette question
sans passer par une modélisation de la décision prise, qui donne un sens au terme ``meilleur".

\subsection{De l'évaluation des procédures par des fonctions de coût et de la dualité avec les
lois a priori}

Il n'est guère difficile de se convaincre qu'il n'existe pas d'estimateur
optimal à l'exception du trivial $\hat{\theta}(x)=\theta$. Suivant les
perspectives adoptées pour classer les estimateurs, certains seront préférables
à d'autres mais très très rarement préférables à tous les autres pour tous les
critères de classement.  C'est du moins le cas dans le cadre fréquentiste où
coexistent en général des classes d'estimateurs faiblement optimaux. A
l'opposé, la perspective bayésienne permet de définir une forme plus forte
d'optimalité et d'aboutir à une solution {\em unique}.

La dérivation de cette propriété repose sur la notion de fonction de perte ou de coût,
empruntée à la théorie des jeux, $L(d,\theta)$, qui évalue l'erreur ou la pénalité résultant
de la décision $d$ lorsque la véritable valeur du paramètre est $\theta$. Par exemple, si
$d$ est la décision d'estimer $\theta$, la formalisation de l'erreur peut être 
la somme ou le maximum des erreurs absolues sur toutes les composantes, soit
$$
L(d,\theta)=\sum_{i=1}^n |\theta_i-d_i|
	\quad\text{ou}\quad
L(d,\theta)=\max_{i=1,\ldots,n} |\theta_i-d_i|\,.
$$
Même si des problèmes véritables peuvent conduire à une détermination unique de
la fonction de coût, dictée par des impératifs financiers par exemple,
il est plus fréquent qu'on doive adopter des fonctions de coût abstraites comme
celles ci-dessus ou la plus traditionnelle, déjà adoptée par Legendre et Laplace,
$$
L_2(d,\theta)=\sum_{i=1}^n (\theta_i-d_i)^2\,,
$$
dont le principal intérêt est de fournir des solutions explicites (voir ci-dessous).

\'Etant donné un problème modélisé par une fonction à valeurs
réelles $L(d,\theta)$, un estimateur $\delta$ propose une décision $\delta(x)$
(ou estimation) pour chaque observation $x$. La comparaison des estimateurs ne
peut se faire au travers de $L(\delta(x),\theta)$, puisque $x$ est aléatoire et
$\theta$ est inconnu. Tandis que l'approche classique évalue les estimateurs au
travers de l'erreur moyenne $\mathbb{E}_\theta[L(\delta(X),\theta)]$ ou risque,
l'approche bayésienne de la théorie de la décision consiste à prendre l'erreur
moyenne par rapport aux paramètres,
$$
\varrho(\delta,x) = \int L(\delta(x),\theta) \pi(\theta|x)\,\text{d}\theta\,.
$$
L'intérêt fondamental de cette perspective est qu'elle permet à la fois d'éliminer
le paramètre inconnu et de conditionner par rapport à l'observation $x$. Les estimations
$\delta(x)$ sont alors toutes comparables à $x$ donné, ce qui permet d'obtenir la meilleure
décision au sens de la fonction de coût : Un {\em estimateur de Bayes} est ainsi 
défini comme la fonction qui associe à tout $x$ la décision
$$
\delta^\pi(x) = \arg\,\min_d\,\int L(d,\theta) \pi(\theta|x)\,\text{d}\theta\,.
$$
Par exemple, le coût $L_2$ ci-dessus conduit à un estimateur de
Bayes égal à la moyenne a posteriori, $\mathbb{E}[\theta|x]$. Cet estimateur classique est
cependant peu recommandé dans le cas de lois a posteriori multimodales, comme dans le cas
des mélanges de distribution \citep{lee:marin:mengersen:robert:2008}.
Sous certaines hypothèses, dont la convexité stricte de la fonction de coût et l'absolue continuité
de la loi a priori, l'estimateur résultant est unique. Il a de plus la propriété de minimiser le
risque intégré,
$$
\int \mathbb{E}_\theta[L(\delta(X),\theta)]\,\text{d}\theta\,,
$$
ce qui lui permet de plus de bénéficier de propriétés fréquentistes d'optimalité comme la
minimaxité--atteindre la plus petite des erreurs maximales-- et l'admissibilité--ne pas connaître
d'autre estimateur dominant $\delta$ uniformément-- \citep[voir][]{robert:2005}. Bien que cette approche de la construction
des procédures statistiques optimales produise des arguments additionnels pour la justification
de l'approche bayésienne (les théorèmes de classes complètes de \citealp{wald:1950} démontrant
ainsi que les seuls estimateurs admissibles sont les estimateurs de Bayes), le recours
à une perspective décisionnelle n'est guère suivi dans les ouvrages récents de statistique
bayésienne (voir par exemple \citealp{gelman:carlin:stern:rubin:2001}). Mon point de vue
est qu'il est quelque peu
regrettable de ne pas prendre en compte les aspects décisionnels, car ils conduisent à une
formulation rigoureuse du choix des estimateurs et illustrent la dualité entre loi a priori
et fonction de coût. En effet, dans la minimisation de
$$
\int L(\delta(x),\theta) \pi(\theta|x)\,\text{d}\theta\,,
$$
seul compte le produit entre fonction de coût et loi a priori. Le transfert de masse entre $L$
et $\pi$ n'a donc aucun effet sur la décision finale. Plus fondamentalement, connaître les
régions de l'espace des paramètres où une erreur est la plus dommageable est similaire à
favoriser dans la loi a priori cette région. C'est d'ailleurs une manière d'aboutir à la
loi de Jeffreys à partir d'une fonction de coût intrinsèque \citep{robert:chopin:rousseau:2009}.

\subsection{De la spécificité des tests en inférence bayésienne, de l'opposition aux $p$-values, et
de deux versions du paradoxe de Lindley}

Le cas particulier des tests statistiques, si souvent utilisés en statistique
classique pour démontrer des ``différences significatives", met en lumière une
distinction majeure entre cette approche classique et l'approche bayésienne,
distinction qui les rend incompatibles sur ce plan. Dans l'approche classique
(des tests), il s'agit d'identifier des événements improbables, c'est à dire
des observations incompatibles avec une certaine loi de probabilité (ou une
famille de lois). Par exemple, dans l'illustration binomiale ci-dessus, on
recherche les valeurs de $X$ les plus improbables lorsque $p=\oh$. Dans cette
approche, l'hypothèse alternative y joue un rôle mineur voire négligeable,
servant au mieux à définir la zone des statistiques extrêmes. L'approche
bayésienne vise à une résolution complète en modélisant les deux hypothèses
simultanément, permettant d'une part la résolution du problème décisionnel du
choix de l'hypothèse la plus probable (choix que certains bayésiens récusent)
et d'autre part la construction d'une loi a posteriori sur les paramètres du
modèle retenu. Le choix d'une hypothèse comme $H_0:\,p=\oh$ se fait donc {\em
contre} ou relativement à une hypothèse alternative comme $H_1:\,p>\oh$ et non
pas dans un absolu mal défini et évitant la modélisation des contraires (et des
décisions à prendre en cas de rejet de l'hypothèse nulle). Cette opposition
entre approche classique et approche bayésienne est souvent mal comprise, d'une
part à cause du caractère relatif associé à la seconde--où deux modélisations
du phénomène sont opposées--et d'autre part du caractère apparemment absolu de
la validation fréquentiste du taux d'erreur de première espèce des tests de
Neyman-Pearson. Un test de Neyman-Pearson au niveau 5\%~a en effet une
probabilité de $0.05$ de rejeter l'hypothèse nulle à tort. Ce taux est en fait
très souvent interprété à tort comme la probabilité de l'hypothèse nulle, ce
qui n'a pas de sens en dehors d'une perspective bayésienne.\footnote{L'approche
bayésienne des tests est souvent qualifiée de ``holmesienne" en référence à la
citation de Sherlock Holmes sur la sélection de l'hypothèse la moins improbable
: {\em ``When you have eliminated the impossible, whatever remains, however
improbable, must be the truth"\/}.} Au delà de l'opposition philosophique, et
des interprétations différentes des résultats, les deux principes conduisent
aussi à des oppositions claires dans les décisions, en particulier parce que
les tests classiques tendent à rejeter plus fréquemment les hypothèses nulles.

La procédure de décision bayésienne est fondée sur la probabilité a posteriori
de l'hypothèse nulle 
$$
\pi(H_0|x) =  \dfrac{\int_{H_0} \pi(\theta) \ell(\theta|x)\,\dd\theta}
{\int \pi(\theta) \ell(\theta|x)\,\dd\theta}\,.
$$ 
ou, de manière (presque) équivalente,\footnote{Le ``presque" est dû au fait
que le facteur de Bayes ne dépend plus des probabilitiés a priori des deux
hypothèses en jeu.} sur le facteur de Bayes \citep{jeffreys:1939}
$$
\mathfrak{B}_{01} (x) = \dfrac{\int_{H_0} \pi(\theta) \ell(\theta|x)\,\dd\theta}
{\int_{H_1} \pi(\theta) \ell(\theta|x)\,\dd\theta} \,\Big/ \dfrac{\int_{H_0} 
\pi(\theta)\,\dd\theta} {\int_{H_1} \pi(\theta) \,\dd\theta} \,.
$$
Ce dernier se compare à $1$ comme la probabilité a posteriori se compare à
$\oh$.  Dans l'exemple binomial, nous avons ainsi vu une valeur de $\mathfrak{B}_{01} (x) 
= 18.95$ qui nous permet de conclure en faveur de $H_0$. (La probabilité a posteriori serait
alors $0.95$.) Dans un cadre classique, la $p$-value est la probabilité de
dépasser $x=117$ sous l'hypothèse nulle, soit $\mathbb{P}(X\ge 117|H_0)=0.42$, qui
s'avère beaucoup moins favorable à $H_0$.

L'opposition mentionnée ci-dessus est des plus claires dans une étude menée par
\cite{berger:sellke:1987}: ils démontrent que la borne inférieure des
probabilités a posteriori est supérieure à la $p$-value (les deux quantités
sont comparables d'un point de vue décisionnel, en utilisant une fonction de
coût 0--1, voir \citealp{robert:2005}, Chapitre 5). Cela signifie que pour {\em
toute} loi a priori donnant le même poids aux deux hypothèses, l'hypothèse
nulle $H_0$ sera {\em toujours} jugée plus improbable par l'approche classique.
\cite{berger:sellke:1987} remarquent par exemple que, en simulant les données
et paramètres suivant une loi a priori spécifique, dans 20\% des cas où
l'hypothèse nulle est rejetée, elle est en fait correcte.

Plus explicitement encore, le paradoxe de \cite{lindley:1957} exprime cette opposition
: dans un modèle normal de moyenne $\theta$ et variance connue avec $n$ observations, lorsque
$H_0:\,\theta=0$, pour une $p$-value donnée, $p(\bar x_n)=\alpha$, sous une loi
a priori normale arbitraire, la probabilité a posteriori de l'hypothèse nulle
tend vers $1$ avec la taille de l'échantillon $n$.  Les interprétations de ce
paradoxe abondent, allant du rejet des $p$-values (car pourquoi faudrait-il garder
cette $p$-value ou le seuil d'acceptation constant?) à celui des probabilités a
posteriori \citep{spanos:2013}, voire les deux
\citep{sprenger:2013}.\footnote{Comme je le souligne dans \cite{robert:2013},
l'opposition est naturelle et doit être relativisée par l'existence de
solutions convergentes dans les deux approches.}

Une autre spécificité des tests bayésiens est de nécessiter une attention particulière
envers les lois impropres, mesures de masse infinie ne pouvant être transformées en lois
de probabilité. D'un point de vue strictement mathématique, ces
lois ne peuvent être employées car elles introduisent une constante de
normalisation arbitraire dans le calcul de la loi a posteriori. Cela conduit à
la seconde forme du paradoxe de \cite{lindley:1957} : étant donné une
observation $x$ arbitraire d'une loi normale, la probabilité a posteriori de
l'hypothèse nulle $H_0:\,\theta=0$ tend vers $1$ quand la variance a priori sous
l'hypothèse alternative tend vers $+\infty$. En effet, faire tendre la variance vers l'infini revient
à utiliser la mesure de Lebesgue comme loi a priori.

Il existe de nombreuses réponses à ce dilemme, de la prohibition des lois impropres
\citep{degroot:1970} à diverses stratégies de validation croisée (ou non-croisée) utilisant
une partie des données pour contruire une véritable loi (a priori ? a posteriori ? a mediori ?) et le reste
des données pour effectuer le test \citep{berger:pericchi:1996}, voire utilisant {\em
toutes} les données pour contruire une loi et les réutilisant une seconde fois pour le test
\citep{aitkin:1991, aitkin:2010, gelman:etal:2013}. Je ne mentionnerai pas ici les alternatives
sur des ``critères d'information" comme le BIC (pour {\em Bayesian Information Criterion}) et
le DIC (pour {\em Deviance Information Criterion}), qui fournissent des classements de modèles
ou d'hypothèses en opposition en dehors des cadres décisionnel {\em et} bayésien.\footnote{Ces
critères sont néanmoins populaires car (i) ne nécessitant pas de loi a priori pour BIC et (ii)
disponibles directement sur le logiciel BUGS pour DIC \citep{lunn:bugs:2012}.}

\section{Envoi}\label{sec:rflX}

Ce bref chapitre n'a pas abordé de nombreux points et modèles où l'approche
bayésienne fait sens, soit pour des raisons de modélisation (comme l'émergence
des modèles graphiques à la fin des années 1980, définis uniquement par des
lois conditionnelles, voir par exemple \citealp{lauritzen:1996}), soit pour des
raisons de complexité générique des modèles (comme dans le cadre de modèles
hiérarchiques rencontrés en marketing comme en sélection animale, en
climatologie comme en épidémiologie), soit encore grâce à sa démarche intégrée
permettant de mêler description stochastique et impératifs de décision (comme
par exemple dans la construction de plans d'expérience dans
\citealp{mueller:parmigiani:robert:rousseau:2004}). Je n'ai pas mentionné non
plus les nouvelles capacités de l'analyse bayésienne à la frontière entre
statistique et machine learning, comme le classificateur BART de
\cite{chipman:george:mcculoch:2006}, techniques qui, associées avec des
méthodologies non-paramétriques aussi rapides, autorisent le traitement de
grands jeux de données, la norme de l'ère du ``Big Data".

Bien que la statistique bayésienne ait principalement grandi dans un terreau
philosophique et polémique, de Bayes et Price s'opposant à Hume, à
\cite{jeffreys:1939} contre Fisher, \cite{keynes:1921} rejettant Laplace
\citep{robert:2010}, \&tc., il est assez paradoxal que l'émergence d'une
composante bayésienne significative dans la statistique (ou science des
données?) moderne soit surtout dû à sa maniabilité face à des modèles complexes
et à l'émergence de techniques de calcul par ordinateur finalement assez
simples à comprendre et implémenter pour que des disciplines utilisatrices de
la statistique s'en saisissent. On peut le regretter du point de vue des
fondements de la discipline, mais la dissémination des idées bayésiennes dans
les autres sciences et au-delà fournit de nouvelles
perspectives et propositions à même d'assurer la continuation de la discipline
dans les décennies à venir : {\em From practice stems theory!}


\end{document}